\documentclass[twocolumn,aps,showpacs,superscriptaddress]{revtex4}
\usepackage{graphicx}% Include figure files
\usepackage{dcolumn}% Align table columns on decimal point
\usepackage{bm}% bold math
\begin{document}
% A useful Journal macro
\def\Journal#1#2#3#4{{#1} {\bf #2}, #3 (#4)}

% Some useful journal names
\def\NCA{Nuovo Cimento}
\def\NIM{Nucl. Instr. Meth.}
\def\NIMA{{Nucl. Instr. Meth.} A}
\def\NPB{{Nucl. Phys.} B}
\def\NPA{{Nucl. Phys.} A}
\def\PLB{{Phys. Lett.}  B}
\def\PRL{Phys. Rev. Lett.}
\def\PRC{{Phys. Rev.} C}
\def\PRD{{Phys. Rev.} D}
\def\ZPC{{Z. Phys.} C}
\def\JPG{{J. Phys.} G}
\def\CPC{Comput. Phys. Commun.}
\def\EPJ{{Eur. Phys. J.} C}

\preprint{}
\title{Identified baryon and meson distributions at large transverse momenta from Au+Au collisions at $\sqrt{s_{_{NN}}} = 200$ GeV}

\affiliation{Argonne National Laboratory, Argonne, Illinois 60439}
\affiliation{University of Birmingham, Birmingham, United Kingdom}
\affiliation{Brookhaven National Laboratory, Upton, New York
11973} \affiliation{California Institute of Technology, Pasadena,
California 91125} \affiliation{University of California, Berkeley,
California 94720} \affiliation{University of California, Davis,
California 95616} \affiliation{University of California, Los
Angeles, California 90095} \affiliation{Carnegie Mellon
University, Pittsburgh, Pennsylvania 15213}
\affiliation{University of Illinois, Chicago}
\affiliation{Creighton University, Omaha, Nebraska 68178}
\affiliation{Nuclear Physics Institute AS CR, 250 68
\v{R}e\v{z}/Prague, Czech Republic} \affiliation{Laboratory for
High Energy (JINR), Dubna, Russia} \affiliation{Particle Physics
Laboratory (JINR), Dubna, Russia} \affiliation{University of
Frankfurt, Frankfurt, Germany} \affiliation{Institute of Physics,
Bhubaneswar 751005, India} \affiliation{Indian Institute of
Technology, Mumbai, India} \affiliation{Indiana University,
Bloomington, Indiana 47408} \affiliation{Institut de Recherches
Subatomiques, Strasbourg, France} \affiliation{University of
Jammu, Jammu 180001, India} \affiliation{Kent State University,
Kent, Ohio 44242} \affiliation{Institute of Modern Physics,
Lanzhou, China} \affiliation{Lawrence Berkeley National
Laboratory, Berkeley, California 94720} \affiliation{Massachusetts
Institute of Technology, Cambridge, MA 02139-4307}
\affiliation{Max-Planck-Institut f\"ur Physik, Munich, Germany}
\affiliation{Michigan State University, East Lansing, Michigan
48824} \affiliation{Moscow Engineering Physics Institute, Moscow
Russia} \affiliation{City College of New York, New York City, New
York 10031} \affiliation{NIKHEF and Utrecht University, Amsterdam,
The Netherlands} \affiliation{Ohio State University, Columbus,
Ohio 43210} \affiliation{Panjab University, Chandigarh 160014,
India} \affiliation{Pennsylvania State University, University
Park, Pennsylvania 16802} \affiliation{Institute of High Energy
Physics, Protvino, Russia} \affiliation{Purdue University, West
Lafayette, Indiana 47907} \affiliation{Pusan National University,
Pusan, Republic of Korea} \affiliation{University of Rajasthan,
Jaipur 302004, India} \affiliation{Rice University, Houston, Texas
77251} \affiliation{Universidade de Sao Paulo, Sao Paulo, Brazil}
\affiliation{University of Science \& Technology of China, Hefei
230026, China} \affiliation{Shanghai Institute of Applied Physics,
Shanghai 201800, China} \affiliation{SUBATECH, Nantes, France}
\affiliation{Texas A\&M University, College Station, Texas 77843}
\affiliation{University of Texas, Austin, Texas 78712}
\affiliation{Tsinghua University, Beijing 100084, China}
\affiliation{Valparaiso University, Valparaiso, Indiana 46383}
\affiliation{Variable Energy Cyclotron Centre, Kolkata 700064,
India} \affiliation{Warsaw University of Technology, Warsaw,
Poland} \affiliation{University of Washington, Seattle, Washington
98195} \affiliation{Wayne State University, Detroit, Michigan
48201} \affiliation{Institute of Particle Physics, CCNU (HZNU),
Wuhan 430079, China} \affiliation{Yale University, New Haven,
Connecticut 06520} \affiliation{University of Zagreb, Zagreb,
HR-10002, Croatia}

\author{B.I.~Abelev}\affiliation{Yale University, New Haven, Connecticut 06520}
\author{M.M.~Aggarwal}\affiliation{Panjab University, Chandigarh 160014, India}
\author{Z.~Ahammed}\affiliation{Variable Energy Cyclotron Centre, Kolkata 700064, India}
\author{B.D.~Anderson}\affiliation{Kent State University, Kent, Ohio 44242}
\author{M.~Anderson}\affiliation{University of California, Davis, California 95616}
\author{D.~Arkhipkin}\affiliation{Particle Physics Laboratory (JINR), Dubna, Russia}
\author{G.S.~Averichev}\affiliation{Laboratory for High Energy (JINR), Dubna, Russia}
\author{Y.~Bai}\affiliation{NIKHEF and Utrecht University, Amsterdam, The Netherlands}
\author{J.~Balewski}\affiliation{Indiana University, Bloomington, Indiana 47408}
\author{O.~Barannikova}\affiliation{University of Illinois, Chicago}
% ...

\collaboration{STAR Collaboration}\noaffiliation

\date{\today}
\begin{abstract}
Transverse momentum spectra of $\pi^{\pm}$, $p$ and $\bar{p}$ up
to 12~GeV/$c$ at mid-rapidity in centrality selected Au+Au
collisions at $\sqrt{s_{_{NN}}} = 200$ GeV are presented. In
central Au+Au collisions, both $\pi^{\pm}$ and $p(\bar{p})$ show
significant suppression with respect to binary scaling at $p_T
\gtrsim $ 4~GeV/$c$. Protons and anti-protons are less suppressed
than $\pi^{\pm}$, in the range 1.5 $\lesssim p_{T} \lesssim$
6~GeV/$c$. The $\pi^-/\pi^+$ and $\bar{p}/p$ ratios show at most a
weak $p_T$ dependence and no significant centrality dependence.
The $p/\pi$ ratios in central Au+Au collisions approach the values
in p+p and d+Au collisions at $p_T \gtrsim$ 5~GeV/$c$. The results
at high $p_T$ indicate that the partonic sources of $\pi^{\pm}$,
$p$ and $\bar{p}$ have similar energy loss when traversing the
nuclear medium.

\end{abstract}
 \maketitle

Ultra-relativistic heavy ion collisions provide a unique
environment to study properties of strongly interacting matter at
high temperature and energy density. When hard partons traverse
the hot and dense medium created in the collision, they lose
energy by gluon radiation and/or colliding elastically with
surrounding partons~\cite{jetquench,starhighpt,rhicotherhighpt}.
This leads to a softening of the hadron spectra at high $p_T$. The
amount of energy loss can be calculated in Quantum Chromodynamics
(QCD) and is expected to be different for energetic gluons, light
quarks and heavy quarks~\cite{xinnian:98,deadcone}. Bulk particle
production at low $p_T$ is dominated by soft QCD processes and the
transverse momentum ($p_T$) distributions are described by
hydrodynamical models incorporating local thermal equilibrium and
collective flow ~\cite{derekhydro,pisahydro,kolbhydro}. Between
these two extreme $p_T$ scales, distinct patterns of meson and
baryon suppression have been observed~\cite{starv2raa,phenixpid},
which are consistent with hadronization through coalescence of
constituent quarks from a collective partonic
system~\cite{voloshin,hwa,fries,ko}.

In this Letter, we present the $p_T$ distributions of pions
($\pi^{\pm}$), protons ($p$) and anti-protons ($\bar{p}$), their
nuclear modification factors, and particle ratios in 200 GeV Au+Au
collisions at 0.3 $< p_{T} <$ 12~GeV/$c$. This explores the full
range of particle production mechanisms, with emphasis on the
intermediate $p_T$ (2 $\lesssim p_T \lesssim $ 6~GeV/$c$) range,
where coalescence may play a role in hadronization, and high $p_T$
($p_T \gtrsim$ 6~GeV/$c$), where particle production is dominated
by jet fragmentation. Identified particles at high $p_T$ provide
direct sensitivity to differences between quark and gluon
fragmentation. For example, proton and pion production at high
$p_T$ is expected to have significant contributions from quark
fragmentation while anti-protons are mostly from gluon
fragmentation~\cite{xinnian:98,AKK}. Therefore, $\bar{p}/p$ and
$\bar{p}/\pi$ ratios in different systems are sensitive to the
possible color charge dependence of energy loss~\cite{xinnian:98}.
We discuss the possible transition between jet fragmentation and
quark coalescence at hadronization, the color charge dependence of
the energy loss, and the fragmentation functions at high $p_T$.

The data used for this analysis were taken in the year 2004 by the
STAR experiment~\cite{STAR}. A total of 15 million central
triggered events for the most central bin (0-12\% total cross
section) and 14 million minimum-bias (MB) triggered events for the
other centrality classes are used~\cite{centralminbias}.
Measurements of the ionization energy loss ($dE/dx$) of charged
tracks in the Time Projection Chamber (TPC) gas are used to
identify pions (protons) in the region $p_{T} \le 0.75$ ($ \le
1.1$)~GeV/$c$ and $2.5  \le p_{T} \le
12$~GeV/$c$~\cite{tpc,pidNIMA}. A prototype Time-of-Flight
detector (TOFr) covering $\pi/30$ rad in azimuth and
$-1\!<\!\eta\!<\!0$ in pseudorapidity~\cite{pidNIMA}, is also
used. By combining the particle identification capability of
$dE/dx$ from the TPC and velocity from the TOFr, pions and protons
can be identified up to 5 GeV/$c$~\cite{pidNIMA,starcronin}. A
detailed description of particle identification throughout the
whole $p_T$ range ($0.3 \le p_{T} \le 12$~GeV/$c$) can be found
in~\cite{pidNIMA}.

At $p_{T} \ge$ 2.5~GeV/$c$, the $dE/dx$ resolution of the TPC is
better than $8\%$ and pions are separated from kaons and protons
on the level of 1.5-3.0 standard deviations in
$dE/dx$~\cite{tpc,pidNIMA}. The prominent yield of the pions can
be extracted from a three-Gaussian fit to the inclusive positively
or negatively charged hadron $dE/dx$ distributions at given
momenta~\cite{pidNIMA,ppdAuPID}. For protons, we used two methods.
One method based on track-by-track selection, using a cut in
$dE/dx$. The other method involved a fit of the $dE/dx$
distribution with three Gaussians~\cite{pidNIMA,ppdAuPID}. For
both methods, the $K_{S}^{0}$ measurement~\cite{starv2raa} is used
to constrain the kaon contribution. The yields presented here are
the results averaged from these two methods.

Acceptance and tracking efficiency are studied by Monte Carlo
GEANT simulations~\cite{starcronin,antiproton}. Weak-decay
feed-down (e.g. $K_{S}^{0}\rightarrow\pi^{+}\pi^{-}$) to the pion
spectra was calculated using the measured $K_{S}^{0}$ and
$\Lambda$ spectra~\cite{starv2raa} and GEANT simulation. The
feed-down contribution was subtracted from the pion spectra and
found to be $\sim12\%$ at $p_{T}=$ 0.35~GeV/$c$, decreasing to
$\sim5\%$ for $p_{T}\gtrsim$ 1~GeV/$c$. Inclusive $p$ and
$\bar{p}$ production is presented without hyperon feed-down
correction in all the figures and discussions. Protons and
anti-protons from hyperon decays have similar detection efficiency
as primordial $p$ and $\bar{p}$ at low $p_T$. At $p_T > $ 2.5
GeV/$c$, the efficiency difference due to decay topology is
estimated to result in a $<$ 10\% correction in final inclusive
yields and is corrected for. The full magnitude of the correction
is assigned as a systematic uncertainty.
\begin{figure}
\includegraphics*[keepaspectratio,scale=0.45]{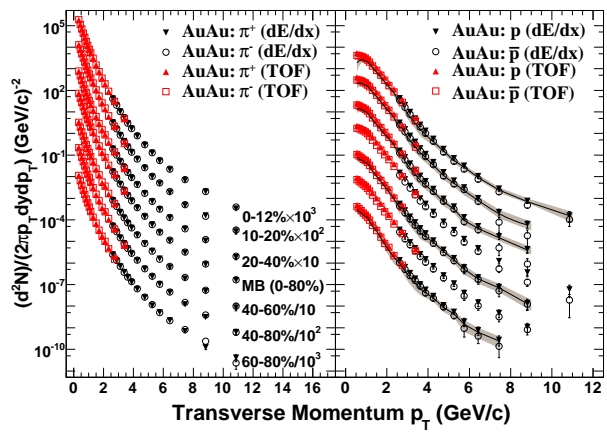}
\vspace{-0.5cm} \caption{Centrality dependence of mid-rapidity
($|y|<0.5$) $\pi^{\pm}$, $p$ and $\bar{p}$ invariant yields versus
$p_{T}$ from 200 GeV Au+Au collisions. The error bars are the
quadrature sum of statistical and systematic errors. The solid
lines depict our best estimates of the proton yields corrected for
the hyperon ($\Lambda$ and $\Sigma^{+}$)
feed-down~\cite{feeddownCorrection}. The shaded bands on the lines
represent the uncertainties. The order of the spectra in different
centralities is the same for both panels.} \vspace{-0.5cm}
\label{Fig:AuAuSpectra}
\end{figure}

\begin{figure}
\includegraphics*[keepaspectratio,scale=0.45]{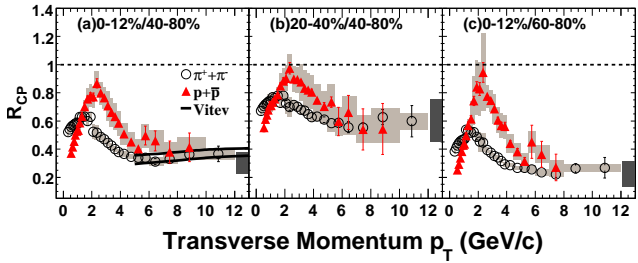}
\vspace{-0.5cm}\caption{Nuclear modification factors $R_{CP}$ for
$\pi^{+}+\pi^{-}$ and $p+\bar{p}$ in 200 GeV Au+Au collisions. The
point-to-point systematic uncertainties are shown as the shaded
boxes around the data points. The dark shaded bands show the
normalization systematic uncertainty in the number of binary
collisions. The solid lines show jet quenching predictions for
pions~\cite{vitevquench}. } \vspace{-0.5cm}\label{Fig:Rcp}
\end{figure}

\begin{figure}
\includegraphics*[keepaspectratio,scale=0.45]{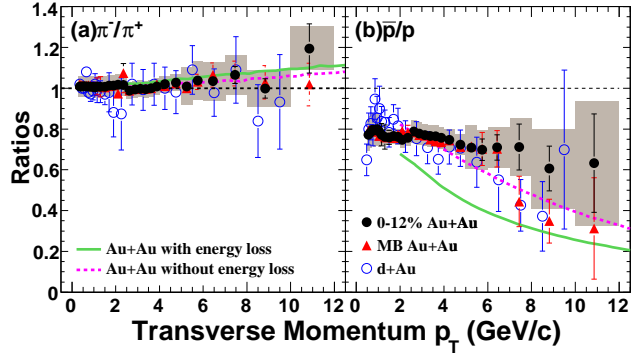}
\hspace{-0.5cm}\vspace{-0.5cm}\caption{The $\pi^{-}/\pi^{+}$ and
$\bar{p}/p$ ratios in 12\% central, MB Au+Au and
d+Au~\cite{starcronin,ppdAuPID} collisions at $\sqrt{s_{_{NN}}} =
200$ GeV. The shaded boxes represent the systematic uncertainties
in the top 12\% central Au+Au collisions. The systematic
uncertainties for MB Au+Au collisions are similar. Curves are the
corresponding predictions from a jet quenching
model~\cite{xinnian:98}.} \vspace{-0.5cm} \label{Fig:antiratio}
\end{figure}

\begin{figure}
\includegraphics*[keepaspectratio,scale=0.45]{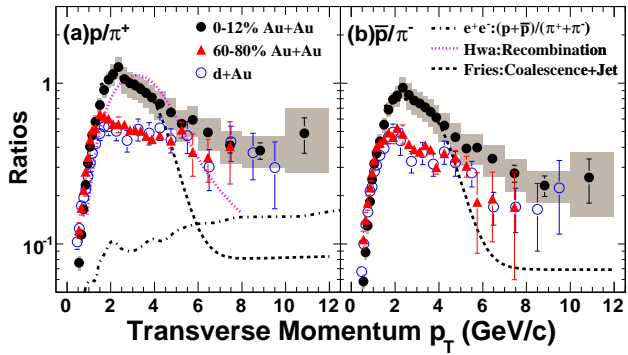}
\hspace{-0.5cm}\vspace{-0.5cm}\caption{The $p/\pi^{+}$ and
$\bar{p}/\pi^{-}$ ratios from d+Au~\cite{starcronin,ppdAuPID} and
Au+Au collisions at $\sqrt{s_{NN}}=200$ GeV. The
$(p+\bar{p})/(\pi^{+}+\pi^{-})$ ratio from light quark jets in
$e^{+}+e^{-}$ collisions at $\sqrt{s}=91.2$ GeV are shown as a
dotted-dashed line~\cite{delphi}. The shaded boxes represent the
systematic uncertainties in the top 12\% central Au+Au collisions.
The systematic uncertainties for 60-80\% Au+Au collisions are
similar. The dotted and dashed lines are model calculations in
central Au+Au collisions~\cite{hwa,fries}.} \vspace{-0.5cm}
\label{Fig:ppirat}
\end{figure}

The invariant yields $d^2N/(2{\pi}p_Tdp_Tdy)$ of $\pi^{\pm}$, $p$
and $\bar{p}$ from Au+Au collisions are shown in
Fig.~\ref{Fig:AuAuSpectra}. The lines in the figure show the
proton spectra after feed-down correction, to illustrate the size
of the estimated feed-down
contribution~\cite{antiproton,pdg,feeddownCorrection}. Systematic
errors for the TOFr measurements are around 8\% and a detailed
list of contributions can be found in previous
publications~\cite{starcronin,Lijuan:04}. Systematic errors for
the TPC measurements are $p_T$ dependent and include uncertainties
in efficiency ($\sim$ 7\%), $dE/dx$ position and width (10-20\%),
$K_{S}^{0}$ constraint (5\%), background from decay feed-down and
ghost tracks (8-14\%), momentum distortion due to charge build-up
in the TPC volume (0-10\%), the distortion of the measured spectra
due to momentum resolution (0-5\%) and half of the difference
between the two methods to extract the proton yields (3-6\%). The
systematic errors are added in quadrature. The spectra from the
TOFr and TPC measurements agree within systematic errors in the
overlapping $p_T$ region. The correlations of the systematic
errors on the particle ratios in Fig.~\ref{Fig:Rcp},
~\ref{Fig:antiratio} and ~\ref{Fig:ppirat} are properly taken into
account.

Nuclear effects on hadron production in Au+Au collisions are
quantified through comparison of the spectrum in central Au+Au
collisions to 40-80\%  or 60-80\% peripheral Au+Au collisions,
scaled by the number of underlying binary nucleon-nucleon
inelastic collisions ($N_{bin}$) calculated from a Glauber
model~\cite{starhighpt}, using the ratio
\[R_{CP}=\frac{d^{2}N/(2{\pi}p_{T}dp_{T}dy)(central)/N_{bin}(central)}{d^{2}N/(2{\pi}p_{T}dp_{T}dy)(peripheral)/N_{bin}(peripheral)}  .\]
Fig.~\ref{Fig:Rcp} shows pion ($\pi^{+}+\pi^{-}$) and proton
($p+\bar{p}$) $R_{CP}$ for Au+Au collisions. In 0-12\% central
Au+Au collisions, the pion yield shows strong suppression with
$R_{CP}$ between 0.2 and 0.4 at $p_{T} \gtrsim $ 3~GeV/$c$. This
is consistent with the jet quenching calculation shown in
Fig.~\ref{Fig:Rcp} (a)~\cite{vitevquench}. For each centrality,
the $R_{CP}$ values for protons peak at $p_T \sim$ 2-3~GeV/$c$. At
intermediate $p_T$, $p$ and $\bar{p}$ are less suppressed, with
respect to binary scaling, than $\pi^{\pm}$, but a significant
suppression is still observed in central Au+Au collisions. This is
in contrast to nuclear modification factors in d+Au collisions,
where a significant enhancement is seen for
protons~\cite{ppdAuPID}. Previous measurements at lower transverse
momentum~\cite{phenixpid} showed that $R_{CP}$ for protons is
close to 1 for 1.5 $< p_{T} <$ 4.5 GeV/$c$. Our results agree with
those measurements within systematic errors, but our data do not
suggest that $R_{CP}$ is constant over the range 1.5 $< p_{T} <$
4.5 GeV/$c$ and the extended $p_T$ reach shows that $R_{CP}$ for
protons decreases again at higher $p_{T}$.

The results in Fig.~\ref{Fig:Rcp} clearly show different $R_{CP}$
for protons and pions at intermediate $p_T$. A similar effect has
been observed for $K_{S}^{0}$ and $\Lambda$~\cite{starv2raa}, with
$K_{S}^{0}$ ($\Lambda$) $R_{CP}$ similar to pion (proton)
$R_{CP}$. The grouping of particle production according to the
number of constituent quarks has been attributed to quark
coalescence at hadronization from a collective partonic
medium~\cite{voloshin,hwa,fries,ko}. Our high statistics
measurements show that these effects disappear at high $p_T$,
where baryons and mesons show a common degree of suppression. This
is consistent with the general expectation that collective and
coalescence effects have a finite $p_T$ reach.

Fig.~\ref{Fig:antiratio} shows the $\pi^{-}/\pi^{+}$ and
$\bar{p}/p$ ratios in 0-12\%, MB Au+Au, and
d+Au~\cite{starcronin,ppdAuPID} collisions. We observe that the
$\pi^{-}/\pi^{+}$ ratios are consistent with unity in d+Au, MB and
central Au+Au collisions. Predictions from a pQCD based model with
and without partonic energy loss are consistent with our
data~\cite{xinnian:98}. The same calculation shows a significant
effect from energy loss on the $\bar{p}/p$ ratio
(Fig.~\ref{Fig:antiratio} (b)), due to the large energy loss of
gluons in the medium. Our measurements, in contrast, show little
centrality dependence of the $\bar{p}/p$ ratio at $p_{T} \lesssim
$ 6 GeV/$c$ and a possible increase of the $\bar{p}/p$ ratio at
higher $p_T$ in central Au+Au collisions compared to d+Au
collisions.

Fig.~\ref{Fig:ppirat} shows the $p/\pi^{+}$ and $\bar{p}/\pi^{-}$
ratios in 0-12\%, 60-80\% Au+Au and
d+Au~\cite{starcronin,ppdAuPID} collisions. The ratios in Au+Au
collisions are observed to be strongly centrality dependent at
intermediate $p_{T}$. In central Au+Au collisions, the $p/\pi^{+}$
and $\bar{p}/\pi^{-}$ ratios peak at $p_T\sim$ 2-3~GeV/$c$ with
values close to unity, decrease with increasing $p_T$, and
approach the ratios in d+Au, p+p and peripheral Au+Au collisions
at $p_{T} \gtrsim$ 5~GeV/$c$. The dotted and dashed lines are
predictions for central Au+Au collisions from
recombination~\cite{hwa} and coalescence with jet quenching and
KKP fragmentation functions~\cite{fries,KKP} respectively. These
models can qualitatively describe the $p(\bar{p})/\pi$ ratio at
intermediate $p_T$ but in general under-predict the results at
high $p_T$.

At high $p_T$, the $p/\pi^{+}$ ratios can be directly compared to
results from quark jet fragmentation as measured in $e^{+}+e^{-}$
collisions by DELPHI~\cite{delphi}, indicated by the dotted-dashed
line in Fig.~\ref{Fig:ppirat} (a). The $p/\pi^{+}$ ratio
measurements in d+Au and Au+Au collisions are higher than in quark
jet fragmentation. This is likely due to a significant
contribution from gluon jets to the proton production, which have
a $(p+\bar{p})/(\pi^{+}+\pi^{-})$ ratio up to two times larger
than quark jets~\cite{delphigluon}. A similar comparison cannot be
made for $\bar{p}$ production (Fig.~\ref{Fig:ppirat} (b)), because
there is a significant imbalance between quark ($q$) and
anti-quark ($\bar{q}$) production at high $p_{T}$ in d+Au and
Au+Au collisions and the fragmentation function of $q$ to
$\bar{p}$ can not be readily derived from $e^{+}+e^{-}$
collisions. It is, however, known from lower beam energies, where
quark fragmentation is dominant, that the $\bar{p}/\pi$ and
$\bar{p}/p$ ratios from quark jets are very small
($<$~0.1)~\cite{ppdAuPID,straub}. The large $\bar{p}/\pi^{-}$
ratio of $\approx$ 0.2 seen in Fig.~\ref{Fig:ppirat} (b) is likely
dominated by gluon fragmentation. This is in agreement with AKK
fragmentation functions~\cite{AKK} which describe the STAR data in
p+p collisions~\cite{ppdAuPID}, showing that gluon fragmentation
contributes to 40\% of pion production at $p_{T} \simeq
10$~GeV/$c$ while more than 80\% of $p + \bar{p}$ are from gluon
fragmentation.

At high $p_{T}$, the nuclear modification factor of protons is
similar to that of pions (Fig.~\ref{Fig:Rcp}) and the $p/\pi^{+}$,
$\bar{p}/\pi^{-}$, and $\bar{p}/p$ ratios in central Au+Au
collisions are similar to those in p+p and d+Au
collisions~\cite{ppdAuPID}. These observations indicate that at
sufficiently high $p_T$, fragmentation in central Au+Au and p+p
events is similar and that there is no evidence of different
energy loss for quarks and gluons in the medium. The theoretical
calculations in Fig.~\ref{Fig:antiratio} show that differences in
radiative energy loss are expected to result in measurable changes
in the $\bar{p}/p$ and $\bar{p}/\pi^{-}$ ratios. Those
calculations, however, do not reproduce the measured $p$ and
$\bar{p}$ spectra in p+p collisions~\cite{ppdAuPID}, indicating
that the fragmentation functions for baryon production are not
well known. The determination of baryon fragmentation functions
from elementary collisions and the expected range of validity of
factorization for baryon production are areas of ongoing
investigation~\cite{ppdAuPID,AKK}. In addition, there is some
uncertainty in the mechanism of energy loss. It has been
postulated that the addition of collisional energy loss to
radiative energy loss may explain the large suppression of leptons
from heavy flavor decays in Au+Au
collisions~\cite{phenixeraa,STARQM05eraa}. The latest
calculations~\cite{magaleda,wicks} including collisional energy
loss and path length fluctuations~\cite{surface} show that the
nuclear modification factor of gluons is still expected to be a
factor of three lower than that of light quarks.

We have reported the transverse momentum spectra of pions and
protons at mid-rapidity from 200 GeV Au+Au collisions up to
12~GeV/$c$. Protons and anti-protons are less suppressed than
pions at intermediate $p_T$. At $p_{T} \gtrsim$ 6~GeV/$c$, both
mesons and baryons are strongly suppressed. However, the relative
particle abundances show no system dependence among p+p, d+Au and
Au+Au collisions. These results indicate that the partonic sources
of $\pi^{\pm}$, $p$ and $\bar{p}$ have similar energy loss when
traversing the nuclear medium. Particle identification at high
$p_T$ provides crucial information and new challenges to the
understanding of energy loss and modified parton fragmentation in
strongly interacting matter.

We thank Dr. M. Djordjevic, R.J. Fries, R.C. Hwa, I. Vitev and
X.N. Wang for valuable discussions and for providing the theory
calculations. We thank the RHIC Operations Group and RCF at BNL,
and the NERSC Center at LBNL for their support. This work was
supported in part by the Offices of NP and HEP within the U.S. DOE
Office of Science; the U.S. NSF; the BMBF of Germany; CNRS/IN2P3,
RA, RPL, and EMN of France; EPSRC of the United Kingdom; FAPESP of
Brazil; the Russian Ministry of Science and Technology; the
Ministry of Education and the NNSFC of China; IRP and GA of the
Czech Republic, FOM of the Netherlands, DAE, DST, and CSIR of the
Government of India; Swiss NSF; the Polish State Committee for
Scientific Research; SRDA of Slovakia, and the Korea Sci. $\&$
Eng. Foundation.


\begin{thebibliography}{9}
%introduction reference
\bibitem{jetquench} M. Gyulassy {\it et al.}, nucl-th/0302077;
A. Kovner {\it et al.}, hep-ph/0304151, Review for: Quark Gluon
Plasma 3, Editors: R.C. Hwa and X.N. Wang, World Scientific,
Singapore.

%\bibitem{eloss} M. Gyulassy {\it et al.}, Nucl. Phys. B \textbf{420}, 583 (1994);
%R. Baier {\it et al.}, Ann. Rev. Nucl. Part. Sci. \textbf{50}, 37
%(2000); X. N. Wang, Phys. Rev. C \textbf{61}, 064910 (2000); E.
%Wang {\it et al.}, Phys. Rev. Lett. \textbf{89}, 162301 (2002).

\bibitem{starhighpt} J. Adams {\it et al.},
\Journal{\PRL}{91}{172302}{2003}.

\bibitem{rhicotherhighpt} S.S. Adler
{\it et al.}, \Journal{\PRL}{91}{072301}{2003}; S.S. Adler {\it et
al.}, \Journal{\PRL}{91}{241803}{2003}; B.B. Back {\it et al.},
\Journal{\PLB}{578}{297}{2004}; I. Arsene {\it et al.},
\Journal{\PRL}{91}{072305}{2003}.

\bibitem{xinnian:98} X.N. Wang, \Journal{\PRC}{58}{2321}{1998};
Q. Wang et al., \Journal{\PRC}{71}{014903}{2005}.

\bibitem{deadcone}Y. Dokshitzer {\it et al.}, \Journal{\PLB}{519}{199}{2001}.


\bibitem{derekhydro} D. Teaney {\it et al.}, nucl-th/0110037;
D. Teaney {\it et al.}, \Journal{\PRL}{86}{4783}{2001}.

\bibitem{pisahydro} P. Huovinen, \Journal{\NPA}{715}{299c}{2003}.
%e-Print Archive: nucl-th/0210024

\bibitem{kolbhydro} P. Kolb {\it et al.}, \Journal{\PRC}{67}{044903}{2003}.

\bibitem{starv2raa} J. Adams {\it et al.},
\Journal{\PRL}{92}{052302}{2004}; J. Adams {\it et al.},
nucl-ex/0601042.

\bibitem{phenixpid} K. Adcox {\it et al.},
\Journal{\PRL}{88}{242301}{2002}; S.S. Adler {\it et al.},
\Journal{\PRL}{91}{172301}{2003}.

%\bibitem{delphi} P. Abreu {\it et al.} (DELPHI Collaboration),  Euro. Phys. J. C 17 207, 2000.

%\bibitem{matt} M. Lamont (STAR Collaboration), J.Phys.G30:S963-S968, 2004.%nucl-ex/0403059

%\bibitem{mgpQCD} M. Gyulassy {\it  et al.}, Phys. Rev. Lett. 86 2537, 2001.

\bibitem{voloshin} D. Molnar {\it et al.}, \Journal{\PRL}{91}{092301}{2003}.

\bibitem{hwa} R.C. Hwa {\it et al.},
\Journal{\PRC}{70}{024905}{2004}.

\bibitem{fries}R.J. Fries {\it et al.},
\Journal{\PRC}{68}{044902}{2003}.

\bibitem{ko} V. Greco {\it et al.},  \Journal{\PRL}{90}{202302}{2003}.

\bibitem{AKK} S. Albino {\it et al.}, Nucl. Phys. B \textbf{725}, 181
(2005). Fragmentation functions from this parameterization are
called AKK.


%Technique reference.
\bibitem{STAR}K.H. Ackermann {\it et al.},
\Journal{\NIMA}{499}{624}{2003}.

\bibitem{centralminbias} Centrality
tagging follows Ref.~\cite{centralitytpc}. The central trigger
selected the most central 12\% of the total hadronic cross section
based on an on-line cut of energy deposited in the Zero-Degree
Calorimeters. The pion spectra from 0-10\% MB events and 0-12\%
central events had a 5\% difference in overall scale due to the
different centrality selections.

\bibitem{centralitytpc} C. Adler {\it et al.},
\Journal{\PRL}{89}{202301}{2002}.

\bibitem{tpc} M. Anderson {\it et al.},
\Journal{\NIMA}{499}{659}{2003}.

\bibitem{pidNIMA}M. Shao {\it et al.}, \Journal{\NIMA}{558}{419}{2006}.

\bibitem{starcronin} J. Adams
{\it et al.}, \Journal{\PLB}{616}{8}{2005}.

\bibitem{ppdAuPID} J. Adams {\it et al.},
\Journal{\PLB}{637}{161}{2006}.

\bibitem{antiproton} C. Adler {\it et al.},
\Journal{\PRL}{87}{262302}{2001}; J. Adams {\it et al.},
\Journal{\PRL}{92}{112301}{2004}.

\bibitem{pdg}S. Eidelman {\it et al.},
\Journal{\PLB}{592}{1}{2004}.

\bibitem{feeddownCorrection} The feed-down corrections were estimated using the $\Lambda$
spectra from Ref.~\cite{starv2raa} with a full simulation of
decay, detection efficiency, and momentum resolution. The measured
$\Lambda$ spectra were extrapolated to high $p_T$ assuming
$\Lambda/p=$ 0.2 at $p_{T}=$ 10 GeV/$c$. The $\Sigma^{+}/\Lambda$
ratio was assumed to be 0.35~\cite{pdg}, independent of $p_{T}$.
The systematic uncertainty on the correction was calculated from
the statistical and systematic uncertainties on the inclusive
proton and $\Lambda$ measurements, with a 30\% uncertainty
assigned to the extrapolated $\Lambda$ spectra. An additional 20\%
uncertainty was assigned to account for the uncertainty in the
$\Sigma^{+}$ yields.

\bibitem{Lijuan:04}L. Ruan, Ph.D. thesis,
University of Science and Technology of China, 2004,
nucl-ex/0503018.

\bibitem{vitevquench} I. Vitev, hep-ph/0603010, curves are calculations
with initial gluon rapidity density 1150 in 0-10\% Au+Au and
between 100 and 150 in 40-80\% Au+Au collisions.

\bibitem{KKP} B.A. Kniehl {\it et al.},
\Journal{\NPB}{597}{337}{2001}. Fragmentation functions from this
parameterization are called KKP.

\bibitem{delphi} P. Abreu {\it et al.}, Eur.
Phy. J. C \textbf{5}, 585 (1998).

\bibitem{delphigluon} P. Abreu {\it et al.}, Eur.
Phy. J. C \textbf{17}, 207 (2000).
%        (arXiv:hep-ph/0502188).

\bibitem{straub} P.B. Straub {\it et al.}, \Journal{\PRD}{45}{3030}{1992}.

%\bibitem{collisionaleloss} M. Mustafa,
%\Journal{\PRC}{72}{014905}{2005}; M. Mustafa, M. Thoma, Acta Phys.
%Hung. A 22, 93 (2005).


\bibitem{phenixeraa} S.S. Adler {\it et
al.}, \Journal{\PRL}{96}{032301}{2006}.

\bibitem{STARQM05eraa} B.I. Abelev {\it et al.}, nucl-ex/0607012.

\bibitem{magaleda} M. Djordjevic {\it et al.}, \Journal{\PLB}{632}{81}{2006}.
\bibitem{wicks} S. Wicks {\it et al.}, nucl-th/0512076.

\bibitem{surface} A. Dainese {\it et al.}, Eur. Phys. J. C \textbf{38},
461 (2005); K. Eskola et al., \Journal{\NPA}{747}{511}{2005}.


%\bibitem{junction}I. Vitev {\it et al.}, \Journal{\PRC}{65}{041902}{2002}.
 \end{thebibliography}
\end{document}